# Introducing a New Evaluation Criteria for EMD-Base Steganography Method

Hanieh Rafiei[1], Mojtaba Mahdavi[2], AhmadReza NaghshNilchi[3]

[1]Department of Information Technology, University of Isfahan, Isfahan, Iran, hanieh_rafiee@eng.ui.ac.ir
[2]Department of Information Technology, Faculty of Computer Engineering, University of Isfahan, Isfahan, Iran, m.mahdavi@eng.ui.ac.ir
[3]Department of Artificial Intelligence, Faculty of Computer Engineering, University of Isfahan, Isfahan, Iran, nilchi@eng.ui.ac.ir

**Abstract**
Steganography is a technique to hide the presence of secret communication. When one of the communication elements is under the influence of the enemy, it can be used. The main measure to evaluate steganography methods in a certain capacity is security. Therefore, in a certain capacity, reducing the amount of changes in the cover media, creates a higher embedding efficiency and thus more security of an steganography method. Mostly, security and capacity are in conflict with each other, the increase of one lead to the decrease of the other. The presence of a single criterion that represents security and capacity at the same time be useful in comparing steganography methods. EMD and the relevant methods are a group of steganography techniques, which optimize the amount of changes resulting from embedding (security). The present paper is aimed to provide an evaluation criterion for this group of steganography methods. In this study, after a general review and comparison of EMD-based steganography techniques, we present a method to compare them exactly, from the perspective of embedding efficiency. First, a formula is presented to determine the value of embedding efficiency, which indicates the effect of one or more changes on one or more pixels. The results demonstrate that the proposed embedding efficiency formula shows the performance of the methods better when several changes are made on a pixel compared to the existing criteria. In the second step, we have obtained an upper bound, which determines the best efficiency for each certain capacity. Finally, based on the introduced bound, another evaluation criterion for a better comparison of the methods is presented.

## 1. Introduction

In the last few decades, steganography in digital content is a debate that has attracted the attention of many researchers. The purpose of cryptographic methods is to hide a secret message in a cover media (e.g., image, audio, video, etc.), so that the presence of the secret message cannot be detected by anyone other than the receiver of the message. Generally, the goal of steganography is to secure communications from an eavesdropper (a person or a system that intends to eavesdrop). Steganography techniques consist of two processes: embedding and extraction. In the embedding process, data is compressed and encrypted using an embedding algorithm and hidden in a cover media. As a result, at this stage, a stego media is produced which is sent to the recipient. After receiving the stego media, the receiver performs the extraction process. In this step, the receiver extracts the hidden data from the stego media using an extraction algorithm. If the sender uses a key in the embedding phase, the receiver needs it in the extraction phase. Considering the use of the compression and encryption step before the steganography algorithm, the receiver should use decryption and then decompression algorithms on the extracted message to extract the original message.

The first steganography method in the world of digital data is the Least Significant Bit (LSB) replacement technique that uses an image as a secret message carrier. In this method, an image with an 8-bit format is regarded as the cover media, if each bit of the secret message is equal to the least significant bit of each pixel, there is no change, otherwise, the least significant bit of the pixel is replaced by the bit of the secret message. This method can embed a large number of secret message bits in all the pixels of the image, which is called "LSB Filling" in some papers [1]. However, the Chi-square -based statistical analysis can easily detect the embedding percentage of this method. The Chi-square test discovers the embedding percentage by using the neighborhood pixel value in each pixel pair [2]. After the LSB replacement method, the LSBM method[1] was presented [3]. The LSBM method is an adaptive method that modifies the least significant bit by matching. In the LSBM method, if the hidden data is equal to the least significant bit, there is no change, but if it is not equal, the least significant bit is decreased or increased. This method is more secure compared to the LSBF method. Also, in the additive LSBM method, if the bit of the secret message does not match with the least significant bit of the cover image, the LSB is only increased and we do not have a decrease in this model.

---
[1] Least Significant Bit Matching

There are two main indices of security and capacity to evaluate a steganography method. The general index of security consists of two concepts. The first concept, perceptual transparency, means retaining the visual or audial quality of the cover media after embedding. The second concept is the coping method of hiding method against all kinds of attacks and discovery or detection methods. Higher capacity enables the use of a smaller cover object to embed a message with fixed size, or the ability to embed a larger message into a cover object with fixed size. Since these two features are closely related to each other, the increase of each of them causes the decrease of the other, so the important challenge of steganography methods is to create a balance between these two factors.

Based on the extension and frequency of media-based steganography techniques, in this study, we present a solution for a more accurate comparison of the steganography methods on the image. Also, the methods of steganography on the image can be divided into several groups in terms of the embedding process. One of these classifications focuses on the number of pixels selected for the embedding process. This category includes two single-pixel categories (i.e., methods where embedding is performed on only one pixel in each step) and multi-pixel or pixel group (Methods that embedding is done on a number of pixels, which together form a pixel group). The second category also includes three important subcategories as covering code [4], Matrix Embedding [5] and EMD [6]. The main focus of the present study is to provide a comparison instrument on the EMD method and the relevant methods.

**Our contributions**. Our goal is to present a method for more accurate comparison of embedding efficiency and capacity of EMD- based techniques. Thus, we present a new formula for embedding efficiency, which can distinguish between multiple changes of one pixel and one change of several pixels. Also, for the accurate comparison of the method with each other, we determine an upper bound, which indicates the best embedding efficiency in each specific capacity. Finally, based on the presented bound, another evaluation criterion is introduced to compare the methods with each other.

**Paper organization.** The paper is organized as follows. After the introduction section, section II expresses generally three categories of methods that embed on pixel groups of the image. Then, in the section III, the EMD method is introduced in details. Section IV discusses the existing methods on EMD in general. In section V, the introduced methods are analyzed and compared with existing formulas. In section VI, after presenting the new comparison method, the methods are analyzed again and their results are shown. Also, in the other part of this section, the bound concept is presented and in the last part of this section, a new evaluation criterion based on the bound is introduced. Finally, the last section is dedicated to conclusion.

## 2. Introduction of pixel group- based methods

There are three types of famous cryptographic methods based on embedding in the pixel group, which have received much attention by authors compared to other steganography methods. The first group is related to the existing methods in the field of cover codes [4]. The mentioned methods originate from the theory of coding and its similarity to the issue of steganography. In general, a $COV(\rho, N, n)$ cover function in the Covering Code method set is capable of embedding n bits of the secret message in N pixels, so that the maximum changes are equal to ρ. Hence, the relative payload of this method is equal to $\alpha = n/N$, the change rate is $\rho/N$, and its embedding efficiency is $e = \alpha/(\rho/N) = n/\rho$. To evaluate the efficiency of the existing cover codes, Fridrich plotted a graph of embedding efficiency $\rho/N$ to the inverse capacity $1/\alpha$. Figure 1 shows this graph [4]. Actually, the upper bound indicates the highest efficiency of the covering codes.

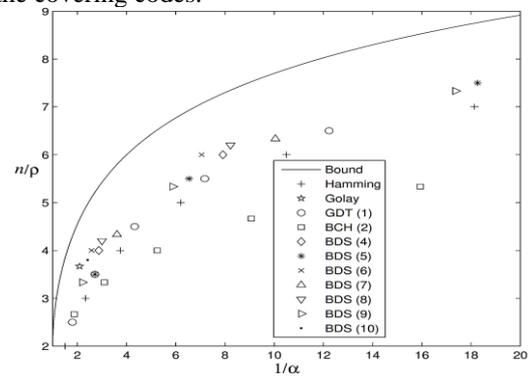

Fig. 1. Embedding efficiency to the inverse capacity for different covering codes [4]

The second category is related to methods based on matrix embedding, which is a type of coding method [5]. This method is presented to increase the efficiency of steganography and reduce the changes caused by embedding secret messages in cover images. Briefly, in methods based on matrix embedding, at first, we divide the image into blocks of a certain length. Then, depending on the matrix embedding method, with a specific mechanism, some of pixels of each block are modified to embed the secret message, with a certain mechanism. For this purpose, most methods based on matrix embedding have a weight matrix.

To increase the embedding efficiency to the inverse capacity in the graph of Fig.1, a third category of steganography methods, called "Exploiting Modification Direction (EMD)" is used [6]. Obtaining the EMD embedding coefficients to increase the embedding payload and efficiency mostly does not pursue a specific process. Normally, the coefficients are obtained using heuristic methods, which is not simple. However, EMD methods generally have more embedding efficiency than covering codes in a certain capacity.

## 3. Exploiting Modification Direction (EMD)

The main idea of this technique is to embed a secret message in a group of $n$ pixels from a cover image [6]. In this method, one pixel is increased or decreased or remains unchanged. Thus, for each group of $n$ pixels of the image, it is possible to change $2n$ different states. In this case, the total number of states that can be used for data embedding is equal to $\mathcal{M} = 2n + 1$, where $2n$ states are for decreasing or increasing $n$ pixels and one state is for when no change is needed. For embedding in a pixel group, $L = \log_2 \mathcal{M}$ bits of the entire secret message $S$ is transferred to the $\mathcal{M}$- Ary notional system and denote it with $s$. If we consider a group of pixel values of the image as a vector $X$, as $X = [x_1, x_2, \ldots, x_n]$, this vector is mapped to a value $f$ in the n-dimensional space that represents the extraction function (1):

$$f(X) = [\sum_{i=1}^{n}(x_i \times i)] \bmod \mathcal{M} \qquad (1)$$

If the secret message ($s$) in the $\mathcal{M}$- Ary notional system is equal to the output of the extraction function ($f$), The pixel group does not change. Else if the secret message ($s$) is not equal to the output of the extraction function ($f$), according to the EMD method, we obtain the value ($d$), which shows the difference between the hidden data value and the current output of the extraction function (2):

$$d = s - f \bmod (2n + 1) \qquad (2)$$

Finally, according to the EMD method, only one pixel is increased or decreased, and according to Equation (3), the changed pixels $G = [g_1, g_2, \ldots, g_n]$ are obtained:

$$\begin{cases} g_d = x_d + 1, g_i = x_i & \text{if } d < f, i \neq d \\ g_{2n+1-d} = x_{2n+1-d} - 1, g_i = x_i & \text{if } d \geq f, i \neq 2n + 1 - d \end{cases} \qquad (3)$$

In this method, only $\mathcal{k} = 1$ pixel in each group may be changed, as the biggest change in that pixel is $z = 1$. Thus, the maximum change in a group will be obtained based on $\rho = z \times \mathcal{k}$. Also, $\alpha = L/n$ bits can be embedded in each pixel relatively, which is called the relative payload. Also, in the EMD method, the embedding efficiency is $E = \alpha/\rho$. Therefore, in the EMD method, for $\alpha = 1.16$ in $n = 2$, $E = 1.16$ embedding efficiency is obtained. All symbols are briefly described in Table 1. Thus, Table 1 is used for the presented symbols [6].

### 4. EMD-based methods

Different recommendations have been presented for the development of the EMD method. For example, instead of changing a pixel in the group or change of one pixel unit, it is possible to make several changes to several pixels at the same time. Hence, we may have more embedding efficiency (E) or embedding payload (α) than EMD. These recommendations have led to the development of the EMD method in various forms. The development of EMD methods can be classified from two aspects. One can be classified as EMD-based methods as the pixel group length is fixed or not. Accordingly, methods 1 to 8 have a fixed pixel group length, and methods 9 to 20 do embedding operations on pixel groups with different lengths. Another view that categorizes EMD-based methods is considering image text during embedding. Except for 20, 8 and 9 methods, which embed the secret message based on the image text, the rest of the methods carry out the embedding operation without considering the image text. Later, the papers are evaluated separately based on the first category. Also, how to extract the secret message in each pixel group is done via the extraction function in all methods, otherwise, the extraction method is included in the method itself.

EMD- based methods, with fixed pixel group

#### 4.1. IEMD

The IEMD method uses two-pixel group $n = 2$ for embedding [7]. Its extraction function is calculated for each pair of pixels $(x_1, x_2)$ as:

$$f(x_1, x_2) = (x_1 \times 1 + x_2 \times 3) \bmod 8 \qquad (4)$$

If the amount of the secret message s is equal to the value of the extraction function $f$ (4), the pair of pixels $(x_1, x_2)$ will remain unchanged, otherwise, it will be turned into one of the following states.

$$\begin{cases} (g_1, g_2) = \\ x_1 = x_1 + 1, x_2 & s = f(x_1 + 1, x_2) \\ x_1 = x_1 - 1, x_2 & s = f(x_1 - 1, x_2) \\ x_1, x_2 = x_2 + 1 & s = f(x_1, x_2 + 1) \\ x_1, x_2 = x_2 - 1 & s = f(x_1, x_2 - 1) \\ x_1 = x_1 + 1, x_2 = x_2 + 1 & s = f(x_1 + 1, x_2 + 1) \\ x_1 = x_1 + 1, x_2 = x_2 - 1 & s = f(x_1 + 1, x_2 - 1) \\ x_1 = x_1 - 1, x_2 = x_2 + 1 & s = f(x_1 - 1, x_2 + 1) \end{cases} \qquad (5)$$

The mentioned method attempts to embed 3 bits in each pair of pixels with a maximum of two changes.

#### 4.2. Pixel Value Adjustment

This method embeds on one pixel of the image. Its extraction function is equal to [8]:

$$\begin{cases} f = (x_i + r) \bmod t^2 \\ r = (x_i - s) \bmod t^2 \end{cases} \quad -\lfloor t^2/2 \rfloor \leq r \leq \lfloor t^2/2 \rfloor \qquad (6)$$

which is $2 \leq t \leq 4$. The r value changes in the specified interval so that $f = s$.

#### 4.3. FEMD

Table 1. Notations and equivalent description

| Notation | Description |
|---|---|
| $I_C$ | Cover Image |
| $I_S$ | Stego Image |
| $H \times W$ | Size of the Image |
| $n$ | Pixel Group Size |
| $f(.)$ | Extraction Function |
| $X = (x_1, x_2, \ldots, x_n)$ | Cover Pixel Group (CPG) |
| $G = (g_1, g_2, \ldots, g_n)$ | Stego Pixel Group |
| $B = (b_1, b_2, \ldots, b_n)$ | Base Vector |
| $\mathcal{M}$ | $\mathcal{M}$ -Ary Notational System (The Number of Symbols) |
| $L = \log_2 \mathcal{M}$ | Length of Secret Message that can be Embed in CPG (bits) |
| $S = \dfrac{H \times W}{n} \times L$ | Maximum Number of Secret Message |
| $s$ | Secret Message that can be embed in CPG |
| $d$ | Difference Between $f$ and $s$ |
| $\mathcal{k}$ | Maximum Number of Changeable Pixels in one Group |
| $z$ | Maximum change of One Pixel |
| $\rho = z \times \mathcal{k}$ | Maximum Number of Change Unit in Each Group |
| $\alpha = L/n$ | Relative Payload (Bit Per Pixel) |
| $E = L/\rho$ | Embedding Efficiency |

This method is presented for $n = 2$ pixel groups [9]. The values of $f$ extraction function in FEMD method is expressed as:

$$f(x_i, x_{i+1}) = [(t-1) \times x_i + t \times x_{i+1}] mod\ t^2 \quad (7)$$

Where, the parameter $t$ is an integer value larger or equal to 2. If the secret message is equal to the output of the extraction function, it remains unchanged. Otherwise, depending on their difference, it uses a change in the range of 1 to $r$ on the pixel group. The value of $r$ is defined as $r = \lfloor t/2 \rfloor$. In brief, FEMD makes effort to embed $L = \log_2 t^2$ bits in each pixel group with maximum $t$ change.

*4.4. DE*

The DE method is presented for the $n = 2$ pixel group. The extraction function of this method is as follows [10]:

$$f(x_1, x_2) = ((2k+1)x_1 + x_2) mod\ 2k^2 + 2k + 1 \quad (8)$$

Were, $k \geq 1$. If the output of the extraction function is not equal to the secret message, one of the members of the neighborhood set of the pixel group replaces it. So, the new pixel group has the least difference with the original group. Generally, this method embeds $L = \log_2 2k^2 + 2k + 1$ bits of the secret message in the pixel group for every $k$ change.

*4.5. APPM*

This method is also for the two-pixel group. Its extraction function is expressed as [11]:

$$f(x_1, x_2) = (x_1 + c_B \times x_2)\ mod\ B \quad (9)$$

The B value is defined based on the size of the image and the total value of the secret message. Also, $c_B$ and the number of changes applied to the pixel group are obtained via the Discrete Optimization Problem. Briefly, the mentioned method attempts to embed $L = \log_2 B$ bits of the secret message in the binary pixel group with minimal change, which relies on the search area of the neighborhood set.

*4.6. Pixel Value Differencing*

The extraction function of this method, which is provided for the binary pixel group, is as follows [12]:

$$f(x_{2i}, x_{2i+1}) = [x_{2i} \times (s_i - 1) + x_{2i+1} \times s_i] mod\ s_i^2 \quad (10)$$

Where $s_i = \log_2 w_i$. Also, this method determines the value of $w_i$ based on the range of the group pixels' difference. The details of the division are fully described in [12]. The number of changes is determined based on the difference of the secret message from the extraction function. Finally, this method embeds $L = \log_2 s_i^2$ bits of the secret message in two pixels.

*4.7. Two-Function*

This method is presented on the binary pixel group [13]. Also, it applies two extraction functions. The general model of its extraction function is defined as:

$$f(x_1, x_2) = (c_1 x_1 + c_2 x_2) mod\ 2^k \quad (11)$$

Where $c_1$ and $c_2$ are relatively prime to each other and $0 < c_1 < c_2 < 2^k$. This method, by Equation (11), defines two extraction functions with different $c_1$, $c_2$, $k$ values. After merging the outputs of the functions with each other, finally, one of its members among the neighborhood set of the pixel group, which has the least changes and is equivalent to the secret message, is selected. The change amount of this method per $L = \log_2 2^{k_1} + \log_2 2^{k_2}$ embedding bit of the secret message in a pixel group depends on its neighborhood set.

*4.8. Kirsch Base*

The amount of embedding in this method is determined based the placement of pixel groups on the edge or outside it [14]. The extraction function of this method on the triple pixel group is as:

$$f = (q_i + 2^t q_{i+1} + 2^z q_{i+2}) mod\ 2^{t+z+1} \quad (12)$$

Where, the values of t and $z$ are determined based on the value of the secret message embedded in each group. Finally, based on the difference between the output of the extraction function and the secret message and the $t$ and $z$ values, possible changes are made in the pixel group. The goal of this method is to embed $L = (t + z + 1)$ secret message bits with a maximum of $3 \times 2^t$ changes in a pixel group.

*4.9. Catalan Base*

In this method, the image is divided into 4-pixel groups and transferred to the Catalan conversion domain [15]. The extraction function of this method is:

$$f = (t_0 + 2^m t_1 + 2^{2m} t_2 + 2^{3m} t_3) mod\ 2^{4m} \quad (13)$$

The $m$ value is determined based on the secret message and capacity, as we have $1\ bpp < m < 4\ bpp$. Finally, changes are applied based on the difference between the output of the extraction function with the secret message and the value of $m$ to the pixel group. Actually, the mentioned method embeds $L = 4m$ bits of the secret message with a maximum of $4 \times 2^{n-1}$ changes in the 4-pixel group.

Methods with non-fixed pixel groups

*4.10. MPEMD*

MPEMD is similar to the EMD method and its extraction function is as follows [16]:

$$f(x_1, x_2, \ldots, x_n) = \left[\sum_{i=1}^{n}(x_i \times i) + C\right] mod\ (2n) \quad (14)$$

Where $n = 2, 3, 4, \ldots$ and $C$ is a security factor, whose value is a random number ranging 0 and $2n - 1$. The method of embedding and the number of changes made on the pixel group is the same as observed in the EMD method. The goal of this method is to embed $L = \log_2 2n$ bits of the secret message with maximum one change in a pixel group.

*4.11. EMD-2*

The extraction function f in this method is calculated using Equation (15) [17]:

$$f(x_1, x_2, \ldots, x_n) = \left[\sum_{i=1}^{n}(x_i \cdot b_i)\right] mod(2w + 1) \quad (15)$$

Where:

$$w = \begin{cases} 4 & n = 2 \\ 8 + 5(n-3) & n > 2, \end{cases} \quad (16)$$

$$[b_1, b_2, \ldots, b_n] = \begin{cases} [1, 3] & n = 2 \\ [1, 2, 6, 11, 16, 21, \ldots, 6 + 5(n-3)] & n > 2 \end{cases} \quad (17)$$

Finally, based on the difference between the output of the extraction function and the secret message and the $w$ value, maximum two changes are applied to the group of $n$ pixels to embed $L = \log_2 2w + 1$ bits of the secret message.

*4.12. 2-EMD*

The 2-EMD method is very similar to EMD [17]. The only difference is the use of $\mathcal{M} = (2n+1)^2$ instead of $\mathcal{M} = (2n+1)$. In this method, $(2n+1)^2$ secret message bits can be embedded in $2n$ pixels consecutively.

*4.13. GEMD*

The GEMD extraction function is as written as [18]:

$$f(x_1, x_2, \ldots, x_n) = \left[\sum_{i=1}^{n} ((2^i - 1) \cdot x_i)\right] \mod 2^{n+1} \quad (18)$$

Depending on the difference between the output of the extraction function and the secret message, a pixel may remain unchanged or a maximum of one unit may change. The goal of this method is to embed $L = n + 1$ bits of the secret message with at most n changes in a pixel group.

*4.14. EGEMD*

GEMD was enhanced in 2017 [19]. The main idea of the method is to divide the pixel group of $n$ $(n > 1)$ into two subgroups and embed $L = n + 2$ bits of the secret message in a pixel group, with a maximum of n changes. For embedding, this method divides the pixel group into two subgroups. Then $L = n + 2$ bits of secret message are decomposed as $s = 2^{n_1+1} \times c + r$, where $r$ and $c$ are two integers and we have $r < 2^{n_1+1}$. Finally, we embed $r$ and c in two-pixel groups using GEMD method, respectively.

*4.15. RGEMD*

That year, another method was presented to increase the capacity of the GEMD method [20]. In this method, first $L = n + 1$ bits of the secret message are embedded in the pixel group using the GEMD method. Then $L' = n - 1$ embeds the next bit of the secret message with a maximum of $n + 1$ changes, in the least significant bit of the pixels embedded by the GEMD method. To extract $n - 1$ bits from the second section of the secret message, it obtains the $LSB(g'_i)$ value. Then, it extracts $n + 1$ bits from the first part of the secret message via GEMD extraction. Finally, the mentioned method embeds $L = 2n$ bits on average in each pixel group with $2n - 1$ changes.

*4.16. Multi-Bit Encoding*

The extraction function of this method is written as [21]:

$$f = [\sum_{i=1}^{n}(b_i \times g_i)] \mod (2^{nk+1}) \quad (19)$$

Where k is an optional value and:

$$b_i = \begin{cases} 1 & i = 1 \\ 2^k b_{i-1} + 1 & i \neq 1 \text{ and } i > 0 \end{cases} \quad (20)$$

Finally, based on the difference between the output of the extraction function and the secret message, with a maximum of $2^{nk+1} - 1$ changes, it embeds $L = nk + 1$ bits of the secret message on the $n$-pixel group [].

*4.17. MSD Base*

The MSD base method uses modified signed digit (MSD) and weight minimization algorithm (WMA) for embedding operation [22] [23]. The extraction function of the MSD base method is as written as:

$$f(x_1, x_2, \ldots, x_n) = \sum_{i=1}^{n} x_i \times 2^{i-1} \mod t_n \quad (21)$$

Where,

$$t_n = \begin{cases} 2 \times \dfrac{4\left\lfloor\frac{n+1}{2}\right\rfloor - 1}{3} + 1, & n \text{ is odd} \\ 4 \times \dfrac{4\left\lfloor\frac{n}{2}\right\rfloor - 1}{3} + 1, & n \text{ is even.} \end{cases} \quad (22)$$

By the WMA algorithm, MSD base minimizes the difference between the output of the extraction function and the secret message. Then it applies the changes based on its binary value on the pixel group. Indeed, based on this method, it is possible to embed $L = \log_2 t_n$ bits in a pixel group with a maximum of n changes, depending on the WMA notation.

*4.18. HEMD*

This method is actually the extension of EMD by multidimensional cube [24]. Its extraction function is as:

$$f(x_1, x_2, \ldots, x_n) = \left(\sum_{i=1}^{n}(x_i \cdot n^{i-1})\right) \mod w^n \quad (23)$$

In the above Equation, $w$ is a parameter that determines that the cube is a multidimensional pixel group and a multi- pixel group. $w$ should be an odd positive number. In fact, this method embeds $\log_2 w^n$ bits in each pixel group, based on the difference between the output of the extraction function and the secret message, with a maximum change of $n(w-1)/2$.

*4.19. EEMDHW*

EEMDHW is an extension of EMD based on Hashed-Weightage Array [25]. Instead of calculating the extraction function, this method calculates the value of $d = mod(x_{index}, 2^k) - s_i$. $x_{index}$ means the pixel in the group is selected by the hashed- weight array. The amount of change per pixel depends on the d value. The mentioned method embeds k bits in each pixel with a maximum of $2^{k-1}$ changes.

*4.20. AEMD*

In this method, before embedding, the edge of the image is detected and then hiding is performed [26]. The edge detection method in this paper is based on Neutrosophic set (MNSED) [27]. It calculates the value of function $f$ using equation (24).

$$f = \left[\sum_{i=1}^{n}(x_i \times b_i)\right] mod\ m^n \quad (24)$$

Where $b = (m^0, m^1, m^2, ..., m^{n-1})$ and $m \geq 2$. Finally, based on the difference between the output of the extraction function and the secret message, this method embeds $L = \lfloor \log_2 m^n \rfloor$ bits of the secret message with a maximum of $n \times \frac{m-1}{2}$ changes in a pixel group.

## 5. Comparison of EMD-based methods based on existing criteria

For a better perception of the presented methods based on EMD, in this section, they are compared based on important and existing parameters. As mentioned in the introduction, two important criteria to evaluate steganography methods are security and capacity. Thus, to measure the capacity criterion, we calculate the relative payload of each method ($\alpha = L/n$). Then, to measure the general measure of security, including the concept of transparency, it is required to calculate the Peak Signal to Noise Ratio (PSNR) for each method. The PSNR parameter, which shows the quality of the embedded image, is obtained from Equation (25). To calculate the value of PSNR, we should obtain the mean square error (MSE) as shown in Equation (26) [28].

$$PSNR = 10 \times \log_{10} \frac{255^2}{MSE} \quad (25)$$

$$MSE = \frac{1}{M \times N} \sum_{x=1}^{M} \sum_{y=1}^{N} (I(x,y) - I'(x,y))^2 \quad (26)$$

Where M and N are the length and width of the image. For better comparison of the methods with each other, first, we bring some of their parameters, which include important information, in Table 2. Information such as pixel group, the number of bits embedded in each group, the maximum change of a pixel, the maximum change of a pixel group and the ability to adapt to the image are also mentioned, which indicates the flexibility of the method to achieve less degradation and the maximum embedding efficiency. The PSNR value of each method is also included. Since each method has stated several PSNR values for different capacities, the numbers that are repeated the most in each paper are reported in Table 2. Apart from PSNR and relative payload, all the information stated in Table 2 is not mentioned directly in the reviewed papers. Therefore, we have obtained this information by the implementation and analysis of the mentioned methods.

The next parameter that is used to evaluate the capacity compared to the changes made in the cover media is the amount of embedding efficiency or embedding efficiency of the method. Actually, embedding efficiency specifies how many bits can be embedded per unit change. In the papers of EMD-based methods, except for the [EMD-ref] method, the amount of embedding efficiency is not mentioned. Also, in the [EMD-Ref] article, the embedding efficiency formula is written as follows.

$$E = \left(\frac{L}{n}\right) / \left(\frac{\rho}{n}\right) = \frac{L}{\rho} \quad (27)$$

Where $\frac{\rho}{n}$ is the relative change in a pixel and $\frac{L}{n}$ is the relative payload. The embedding efficiency formula mentioned in the paper [6] is correct for this method, because there is at most one change in each pixel group and each pixel changes as one at maximum. But it is not accurate for methods whose pixels have more than one change. Table 3 demonstrates the calculation of the embedding efficiency of each method according to the relative payload of the method. These values were not reported in any of the reviewed papers. As mentioned before, we have obtained the value of the embedding efficiency of each method using the formula (27), corresponding to its relative payload by calculating, analyzing and implementing the methods. Table 3 contains a column named "condition". This column includes the conditions that we have considered when analyzing the method to achieve a certain efficiency value.

Table 3. Efficiency with Relative Payload

| Method Name | Condition | Relative Payload ($\alpha$) | Efficiency ($E$) [6] |
|---|---|---|---|
| Catalan Base [15] | - | 1 | 0.125 |
| FEMD [9] | $t = 2$ | 1 | 1 |
| MPEMD [16] | $n = 2$ | 1 | 1 |
| MSD Base [23] | $n = 3$ | 1.15 | 1.15 |
| Kirsch Base [14] | $t = 1, z = 2$ | 1.33 | 0.66 |
| Multi Bit Encoding [21] | $n = 2, k = 1$ | 1.5 | 1 |
| IEMD [7] | - | 1.5 | 1.5 |
| GEMD [18] | $n = 2$ | 1.5 | 1.5 |
| EGEMD [19] | $n = 4$ | 1.5 | 1.5 |
| HEMD [24] | $n = 3, w = 3$ | 1.58 | 1.58 |
| EMD-2 [17] | $n = 2$ | 1.58 | 1.58 |
| DE [10] | $k = 2$ | 1.85 | 1.85 |
| RGEMD [20] | $n = 3$ | 2 | 0.85 |
| Pixel Value Adjustment [8] | $t = 2$ | 2 | 1 |
| AEMD [26] | $m = 4$ | 2 | 1 |
| APPM [11] | $B = 16$ | 2 | 1.33 |
| 2-Function [13] | $k_1 = 2, k_2 = 3$ | 2.5 | 0.71 |
| Pixel Value Differencing [12] | $k_i$ | 2.5 | 0.45 |
| EEMDHW [25] | $k = 4$ | 4 | 0.5 |

Table 2. Introduce Important parameters

| The Method Name & Year of Publish | Number of pixels in a group | Relative Payload ($\alpha$) | Maximum Number of Changeable Pixels in a Group ($k$) | Maximum Change of One Pixel ($z$) | Payload of a Group (L) | PSNR | Adaptivity |
|---|---|---|---|---|---|---|---|
| EMD – 2006 [6] | $n \geq 2$ | $\alpha = \frac{\log_2(2n+1)}{n}$ | 1 | 1 | $L = \log_2 2n + 1$ | 56.15 54.14 | - |
| IEMD – 2007 [7] | 2 | 1.5 | 2 | 1 | 3 | 50.17 | - |
| EMD-2 – 2010 [17] | $n \geq 2$ | $\alpha = (\log_2(2w+1))/n$ | 2 | 1 | $L = \log_2 2w + 1$ | 52.03 49.89 | - |
| DE – 2010 [10] | 2 | $\alpha = \frac{\log_2(2k^2 + 2k + 1)}{2}$ | 2 | $k$ | $L = \log_2(2k^2 + 2k + 1)$ | 52.10 47.80 | - |
| FEMD – 2011 [9] | 2 | $\alpha = \lfloor \log_2 t^2 \rfloor / 2$ | 2 | $2r$ | $L = \lfloor \log_2 t^2 \rfloor$ | 52.39 46.75 | - |
| APPM – 2012 [11] | 2 | $\alpha = (\log_2 B)/2$ | 2 | Related on Search Region | $L = \log_2 B$ | 52.11 47.80 | - |
| GEMD – 2013 [18] | $n$ | $\alpha = (n+1)/n$ | $n$ | 1 | $L = n + 1$ | 50.78 51.02 | - |
| EGEMD – 2017 [19] | $n = n_1 + n_2$ | $\alpha = (n+2)/n$ | $n$ | 1 | $L = n + 2$ | 47.69 47.78 | - |
| RGEMD – 2017 [20] | $n$ | $\alpha = 2$ | $n$ | $n$ | $L = 2n$ | 44.74 | - |
| Pixel Value Differencing – 2015 [12] | 2 | $\alpha = \log_2 s_i^2 / 2$ | 2 | $s_i$ | $L = \log_2 s_i^2$ | 42.46 | - |
| Multi Bit Encoding – 2016 [21] | $n$ | $\alpha = (nk+1)/n$ | $n$ | $2^k - 1$ | $L = nk + 1$ | 50.50 43.00 | - |
| MSD base – 2016 [23] | $n$ | $\alpha = \frac{\log_2 t_n}{n}$ | $n$ | 1 | $L = \log_2 t_n$ | 52.11 51.85 | - |
| HEMD – 2019 [24] | $n \geq 3$ | $\alpha = \frac{\log_2 w^n}{n}$ | $n$ | $(\frac{w-1}{2})$ | $L = \log_2 w^n$ | 49.89 34.33 | - |
| Pixel Value Adjustment – 2020 [8] | 1 | $\alpha = L$ | 1 | $\lfloor t^2/2 \rfloor$ | $L = t$ | 42.84 37.04 | - |
| EEMDHW – 2020 [25] | $n$ | $\alpha = K$ | $n$ | $2^{K-1}$ | $nK$ | 48.52 | - |
| MPEMD – 2020 [16] | $n$ | $\alpha = (\log_2 2n)/n$ | 1 | 1 | $L = \log_2 2n$ | 55.00 53.47 | - |
| Kirsch Base – 2020 [14] | 3 | $\alpha = \frac{t+z+1}{3}$ | 3 | $2^t$ | $L = t + z + 1$ | 44.90 37.35 | + |
| AEMD – 2021 [26] | $n$ | $\alpha = ((\log_2 m_{edg}^n) + (\log_2 m_{non-edg}^n))/n$ | $n$ | $r_L = -\frac{m-1}{2} + 1$ | $L = (\log_2 m_{edg}^n) + (\log_2 m_{non-ed}^n)$ | 46.21 | + |
| Catalan Base -2021 [15] | 4 | $\alpha \in \{1,2,3,4\}$ | 4 | $2^{m-1}$ | $1m$ to $4m$ | 48.62 27.93 | + |
| 2-Function – 2021 [13] | 2 | $\alpha = \frac{\log_2 2^{k_1} + \log_2 2^{k_2}}{2}$ | 2 | Related on Search Region | $L = \log_2 2^{k_1} + \log_2 2^{k_2}$ | 43.72 34.80 | - |

## 6. Introduce the Proposed Measurement Criteria

This section of the study consists of 4 subsections. In the first subsection, we introduce a new criterion for evaluating the embedding efficiency and explain its difference with the criterion introduced in section 5. Also, its value for all methods is calculated. In the second subsection, we present the concept of the upper bound for both criteria (criterion of section 5 and the criterion of section 6.1) and express its relationship. Then, in subsection 3, we estimate an equation for the upper bound with the proposed efficiency formula and obtain the distance of each method from the bound. Finally, in the last subsection, the basis of the calculation of the upper bound is expressed. Thus, when we refer to bound, it means the upper bound.

### 6.1. Introduce the Proposed Criterion of Embedding Efficiency

As discussed in section 5, the embedding efficiency with formula (27) is not accurate for methods whose pixels have more than one change. Thus, we specify how many bits can be embedded per one distortion unit in one pixel. One of the criteria that determines the amount of distortion and indicates transparency is PSNR. Based on the PSNR formula, $\sqrt{MSE}$ can also be used to display the distortion. In fact, $\sqrt{MSE}$ is another expression of distortion and it can be calculated using PSNR value. Thus, our proposed embedding efficiency equation is equal to relative payload (α) to $\sqrt{MSE}$.

$$E' = \frac{\alpha}{\sqrt{MSE}} \quad (28)$$

Equation (29) calculates the value of $\sqrt{MSE}$ in terms of PSNR based on Equation (25). As the value of $\sqrt{MSE}$ is not expressed separately in most papers of EMD-based methods, so we rewrite the Equation of embedding efficiency in terms of $\sqrt{MSE}$ using PSNR, which is dependent on MSE, because in most papers related to these methods, at least one PSNR value has been reported.

$$MSE = \frac{65025}{10^{\frac{PSNR}{10}}} \quad (29)$$

Thus, instead of Equation (28), Equation (30) can be used to calculate $E'$:

$$E' = \frac{\alpha}{\sqrt{\frac{65025}{10^{\frac{PSNR}{10}}}}} \quad (30)$$

For better understanding of the difference between $\rho$ and $\sqrt{MSE}$ in the embedding efficiency equation, consider two EMD-based methods. The first method makes one change in three pixels and the second method makes three changes in one pixel. $\rho$ of two values is 3 for both methods, but the degradation rate of the second method is higher than the first one according to the value of $\sqrt{MSE}$, which will lead to more reduction of the PSNR value of the image. The equation $E = \frac{L}{\rho}$ cannot be a good expression for reducing the given transparency. As the degradation in the image reduces the real embedding efficiency (transparency), the use of $\sqrt{MSE}$ expresses the value of embedding efficiency more accurately. Table 4 calculates and show the proposed embedding efficiency based on the relative payload of each method. Table 4 also includes a column named "condition". The meaning of this column is the conditions that have been considered when analyzing the method to achieve a certain efficiency value.

It is worth to mention that there are differences between the "conditions" and "related payload" columns in Tables 3 & 4. These differences were due to incomplete information in some papers, we attempted to obtain embedding efficiency values based on two criteria under similar conditions. Unfortunately, in some methods, this was not possible. Also, we trusted the reported PSNR values in each method and used one of the best highly repeated values in each method as the basis of our calculations.

### 6.2. Calculation of the Upper Bound of Embedding Efficiency

To have a clear understanding of Table 3, we have plotted a diagram according to positioning efficiency with formula (27) to the inverse capacity according to Figure 2. We have also drawn Figure 3 for Table 4. Figure 3 indicates a graph of efficiency with formula (28) to the inverse capacity. In order to better interpret and compare the methods from the aspects of these two criteria, we need a threshold or a bound. In fact, the bound is a narrow border which specifies the best efficiency for each capacity. Thus, we estimated and considered a bound for each chart. The bound points determine the best embedding efficiency in each related payload. The best efficiency in terms of capacity is also obtained as shown in Formula (27) when the number of changes is the lowest, that is, $argmin(\sum(x - x'))$. Therefore, when we attempt to embed the secret message change with maximum z, when we claim we have reached the best efficiency, when we have first obtained all states of $x - x'$ equal to one. Then, if $z > 1$, we should create all the states where $x - x'$ is equal to 2. After finishing these steps, we take the states where $x - x'$ is equal to 3. In other words, it is possible to have states with a change in $z$ when, before that, we have considered all the states with a change of 1 to $z - 1$ for all pixels. For this purpose, we define the variable q that indicates this concept, that is, the maximum number of pixels that can have $z$ changes, assuming that we have made changes from 1 to $z - 1$ for all pixels before it.

Table 4. Proposed Efficiency with Relative Payload

| Method Name | Condition | Relative Payload ($\alpha$) | Proposed Efficiency ($E'$) |
|---|---|---|---|
| EMD [6] | $n = 3$ | 0.9357 | 1.0107 |
| Catalan Base [15] | - | 1 | 1.057 |
| MPEMD [16] | $n = 2$ | 1 | 1.15 |
| FEMD [9] | $t = 2$ | 1 | 1.6329 |
| MSD base [23] | $n = 3$ | 1.15 | 1.77 |
| DE [10] | $k = 1$ | 1.16 | 1.83 |
| GEMD [18] | $n = 3$ | 1.33 | 1.804 |
| Multi Bit Encoding [21] | $n = 3, k = 1$ | 1.33 | 1.751 |
| IEMD [7] | - | 1.5 | 1.897845 |
| APPM [11] | $B = 9$ | 1.5 | 1.9 |
| Pixel Value Differencing [12] | - | 1.53 | 0.7964 |
| EMD-2 [17] | $n = 2$ | 1.58 | 1.94 |
| HEMD [24] | $n = 3, w = 3$ | 1.58 | 1.94 |
| EGEMD [19] | $n = 3$ | 1.67 | 1.587 |
| Kirsch Base [14] | - | 1.84 | 1.2792 |
| Pixel Value Adjustment [8] | $t = 2$ | 2 | 0.5914 |
| RGEMD [20] | $n = 3$ | 2 | 1.35 |
| AEMD [26] | - | 2.037 | 1.6278 |
| 2-Function [13] | $k_1 = 2, k_2 = 3$ | 2.5 | 1.05 |
| EEMDHW [25] | $k = 2$ | 3 | 1.11 |

Regarding the graph bound with the proposed efficiency formula (28), we reach the best efficiency for each capacity when the MSE value is minimized (when MSE is minimal when we have $argmin(\sum(x - x')^2)$). Thus, we start from the states where $(x - x')^2$ is equal to 1. Then we obtain the states where $(x - x')^2$ is equal to 4. We continue this trend as mentioned above. Therefore, until we have not considered all the states from 1 to $(z-1)^2$ changes, we cannot have states with $z^2$ change for one or more pixels.

So, to estimate and plot each bound, a regression relationship is required to calculate the total number of states and the amount of state changes for different $n$, $q$, and $z$. The bound of the first Diagram is drawn using Equation (31) for different n, k and $z$.

$$E = \frac{log_2 f_\mathcal{M}(n,z,q)}{f_\rho(n,z,q)} \quad (31)$$

Where $f_\mathcal{M}(n,z,q)$ is a recurrence relation defined as follows:

$$f_\mathcal{M}(n,z,q) = \begin{cases} 2f_\mathcal{M}(n-1,z,q-1) + (2z-1).f_\mathcal{M}(n-1,z,q) & n \neq q \\ (2z+1)^n & n = q \\ f_\mathcal{M}(n,z-1,n) = (2z-1)^n & q = 0 \end{cases} \quad (32)$$

As shown in (33), we have also created $f_\rho(n,z,q)$ as a recurrence relation.

$$f_\rho(n,z,q) = \begin{cases} \overline{f_\rho}(n,z-1) + \sum_{i=1}^{q} f_\rho'(n,z,i) & n \neq q \\ \overline{f_\rho}(n,z) & n = q \end{cases} \quad (33)$$

So, we have:

$$f_\rho'(n,z,q) = 2^q \binom{n}{q} . [q.z.\overline{f_\mathcal{M}}(n-q,z-1) + \overline{f_\rho}(n-q,z-1)] \quad (34)$$

And:

$$\overline{f_\rho}(n,z) = \overline{f_\mathcal{M}}(q,z).\overline{f_\rho}(n-q,z) + \overline{f_\mathcal{M}}(n-q,z).\overline{f_\rho}(q,z) \quad (35)$$

In which $\overline{f_\mathcal{M}}(.)$ is defined as follows:

$$\overline{f_\mathcal{M}}(n,z) = (2z+1)^n \quad (36)$$

Also, as the value of $q$ or $z$ is equal to zero, we have:

$$\overline{f_\rho}(n,0) = 0 \quad (37)$$

And,

$$f_\rho(n,z,0) = \overline{f_\rho}(n,z-1) \quad (38)$$

Finally, this recurrence relation continues until the length of the group is equal to one, then we have:

$$\overline{f_\rho}(1,z) = \sum_{i=-z}^{z} i = 2\sum_{i=0}^{z} i \quad (39)$$

The bound of the second diagram for the proposed Equation (28) is drawn using the Equation (40) for different $n$, $q$ and $z$:

$$E' = \frac{\frac{log_2 f_\mathcal{M}}{n}}{\sqrt{f_\rho}} \quad (40)$$

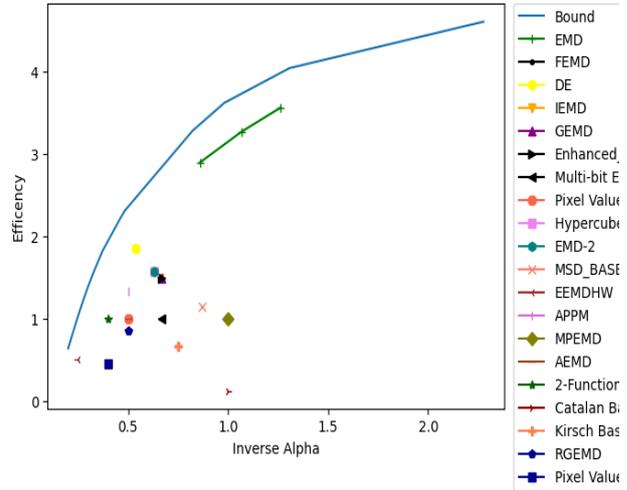

Fig. 2. Comparison of standard embedding efficiency to inverse capacity of EMD- based methods

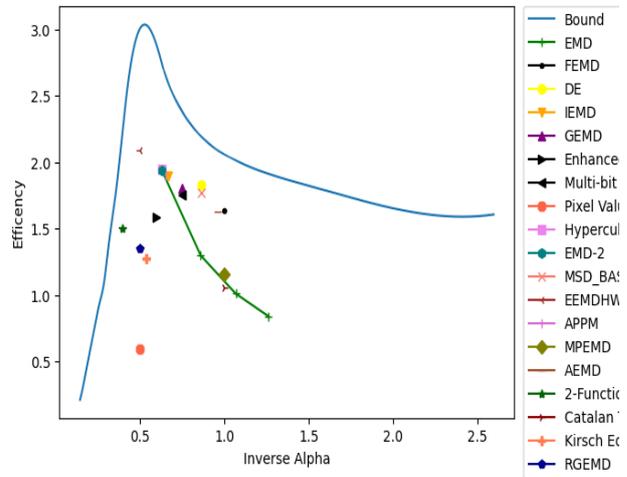

Fig. 3. Comparison of the proposed embedding efficiency to the inverse capacity of the EMD-based methods

Where $f_\mathcal{M}$ is obtained using Equation (32) and $f_\rho$ is obtained as Equation (30). Only the Equation $\overline{f_\rho}(1,z)$ (35) and $f_\rho'(n,z,q)$ (39) changes to Equations (41) and (42).

$$\overline{f_\rho}(1,z) = 2\sum_{i=0}^{z} i^2 \quad (41)$$

$$f_\rho'(n,z,q) = 2^q \binom{n}{q} . [q.z^2.\overline{f_\mathcal{M}}(n-q,z-1) + \overline{f_\rho}(n-q,z-1)] \quad (42)$$

The mathematical basis of recurrence relations is described in details in the subsection 6-4 due to observing the coherence of relations.

### 6.3. Estimation of Upper Bound Equation and Introduction of the Relevant Evaluation Criteria

By the obtained bounds, a criterion can be obtained to compare the methods proposed in references [6-26]. To compare two stenography methods, it is required to obtain the distance of the efficiency point of that method in a certain capacity from the bound curve. As most methods have more than one change in one pixel, we choose the bound based on the proposed formula. The best method is to have the least distance from the bound. To calculate the distance of each method from the bound, the bound can be estimated as a set of lines and the distance of each method can be obtained from these lines. In some cases, the calculation error can be high. It is better to obtain a mathematical relation with at least 3 degrees of freedom, which best matches with the obtained bound points. Among the types of linear and non-linear equations with minimum degrees of freedom, we concluded that the bound points on the Equation (43) of the polynomial type can be matched well.

$$y = 2.994\,x^3 - 10.5\,x^2 + 10.82\,x - 1.098 \quad (43)$$

There may be better equations for matching the bound points; however, considering the implementations and the minimal degree of freedom, we have obtained the best matching using Equation (43). Figure 14 demonstrates the bound Diagram and equation Diagram (43) next to each other. We have also obtained the distance of each method from the bound using the defined relation (43). Thus, all the methods explained in section 4 are written in Table 5 with the value of their distance from the bound. In a certain capacity, the shorter the distance of the method from the bound, the better the method is in terms of efficiency.

Table 5. Distance of Method from The Bound (Fig. 5)

| Method Name | Relative Payload ($\alpha$) | Distance from Bound |
|---|---|---|
| EMD [6] | 0.9357 | 0.3595 |
| FEMD [9] | 1 | 0.5871 |
| MPEMD [16] | 1 | 1.0653 |
| Catalan Base [15] | 1 | 1.1621 |
| MSD Base [23] | 1.15 | 0.5728 |
| DE [10] | 1.16 | 0.5149 |
| GEMD [18] | 1.33 | 0.5708 |
| Multi Bit Encoding [21] | 1.33 | 0.6234 |
| IEMD [7] | 1.5 | 0.4362 |
| APPM [11] | 1.5 | 0.4258 |
| Pixel Value Differencing [12] | 1.53 | 1.5275 |
| EMD-2 [17] | 1.58 | 0.3595 |
| HEMD [24] | 1.59 | 0.3492 |
| EGEMD [19] | 1.67 | 0.67121 |
| Kirsch Base [14] | 1.849 | 0.8758 |
| RGEMD [20] | 2 | 0.7096 |
| Pixel Value Adjustment [8] | 2 | 1.4687 |
| AEMD [26] | 2.037 | 0.6385 |
| 2-Function [13] | 2.5 | 0.2354 |
| EEMDHW [25] | 3 | 0.3395 |

*6.4. The Basis to Draw the Upper Bound*

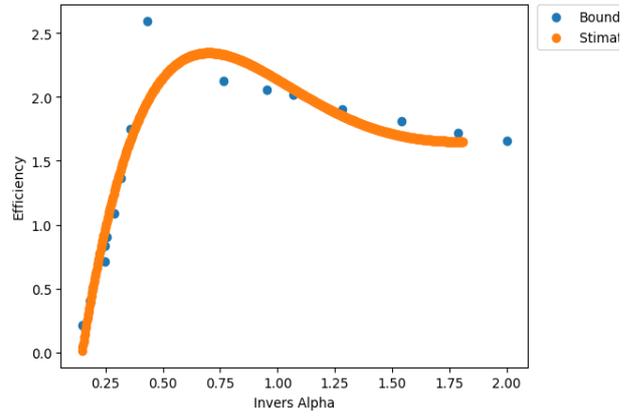

Fig. 5. Bound points next to Equation points (40)

To draw the bounds, we determine two values of all states ($\mathcal{M}$) and the rate of change (ρ) should be determined for a specific pixel group n and the number of selectable pixels q for the maximum change z. Hence, we decided to create relations that receive three specific values (n, z, q) as input and generate $\mathcal{M}$ and ρ values as output. In this regard, we have defined two recurrence relations to achieve each value of $\mathcal{M}$, ρ.

Based on the nature of the variable $q$ which was defined in section 6 to achieve the bound, in order to produce the best value of $\mathcal{M}$, ρ per input $(n, z, q)$, we follow these steps: First, one of the states in which a pixel can be changed is selected. After finishing the one-change states for all pixels, we go to the two-change states. This process continues until we write all states for $\pm(z - 1)$ change. Then we consider the states that include one pixel with $\pm(z)$ change and another $n - 1$ pixel with changes in the range of $-(z - 1)$ to $+(z - 1)$. Finally, we continue the same method until q pixels with $\pm(z)$ changes. As it was explained, the presence of a recurrence relation is necessary to produce the best value of $\mathcal{M}$ and ρ for $(n, z, q)$. Later, how to obtain the recurrence relation related to all states $(f_\mathcal{M}(n, z, q))$ and the recurrence relation related to the number of changes $(f_\rho(n, z, q))$ for the standard embedding efficiency formula and then the proposed embedding efficiency formula are expressed respectively.

*6.4.1. Recurrence Relation Representing All States*

The general form of the recurrence relation as specified for $(n, z, q)$, producing all states is as follows:

$$f_\mathcal{M}(n, z, q) = \begin{cases} 2f_\mathcal{M}(n-1, z, q-1) + (2z-1).f_\mathcal{M}(n-1, z, q) & n \neq q \\ (2z+1)^n & n = q \\ f_\mathcal{M}(n, z-1, n) = (2z-1)^n & q = 0 \end{cases} \quad (44)$$

As shown in Equation (44), three forms are considered to create it. The first is when we have $q \neq n$, thus, at first, a pixel from this group is considered. This pixel either has $\pm z$ change or no $\pm z$ change. If the specified pixel has $\pm z$ change, then we have $n - 1$ pixels from the group, $q - 1$ of which can have $\pm z$ change. Since the specified pixel can have $\pm z$ change, by eliminating it from the group, $2f_\mathcal{M}(n-1, z, q-1)$ states can be generated. The value 2 is the + or - change of $z$, which creates two states for the relevant pixel. Now, if that specific pixel does not change $\pm z$, then it can have the range of changes from $-(z-1)$ to $(z-1)$, it means, $2(z-1)$ states can take place for that pixel. Also, if the $\pm z$ pixel does not change, it can have an unchanged state, so the number of states for that pixel is as follows:

$$2(z-1) + 1 = 2z - 1 \quad (45)$$

Therefore, assuming $2z - 1$ states for a pixel that has no $\pm z$ change, we have $n - 1$ pixels that q of it has $\pm z$ change. The total number of states is equal to $(2z - 1). f_\mathcal{M}(n - 1, z, q)$. Therefore, the total number of states is equal to:

$$f_\mathcal{M}(n, z, q) = 2f_\mathcal{M}(n - 1, z, q - 1) + (2z - 1). f_\mathcal{M}(n - 1, z, q) \quad (46)$$

The first section of the sum equation is when the separated pixel has $\pm z$ change and the second section is when the mentioned pixel has no $\pm z$ changes. The second form is when we have, $q = n$, it means that all pixels can have $\pm z$ change, so each pixel can have $2z$ changes. In addition, each pixel can remain unchanged. Therefore, the number of states for each pixel is $2z + 1$, and since we have $n$ pixels, the total number of states under these conditions is equal to:

$$f_\mathcal{M}(n, z, n) = (2z + 1)^n \quad (47)$$

The third form is when $q = 0$, in this case all pixels can have $\pm(z - 1)$ changes, so the total number of states in this condition is as:

$$f_\mathcal{M}(n, z, 0) = f_\mathcal{M}(n, z - 1, n) = (2z - 1)^n \quad (48)$$

*6.4.2 Recurrence Relationship Representing the Amount of Changes*

In this section, at first how to calculate the sum of changes is considered using an example and then the sum of changes for $n, z,$ and $q$ is obtained.

Example. If $n = 2$, $z = 1$ and $q = 2$, then the states of Table 6 are presented as:

Table 6: The created states for $n = 2$, $z = 1$ and $q = 2$

| State | | Sum of Number |
|---|---|---|
| $-1$ | $-1$ | 2 |
| $-1$ | 0 | 1 |
| 0 | $-1$ | 1 |
| 0 | 0 | 0 |
| 0 | 1 | 1 |
| 1 | 0 | 1 |
| 1 | 1 | 2 |

Therefore, we have $\rho = 8$ and $\mathcal{M} = 7$.
As shown, calculating the sum of changes is more complicated than calculating the total number of states due to the sum of the changes of each state. The equation calculating the sum of the number of changes is as:

$$f_\rho(n, z, q) = \begin{cases} \overline{f_\rho}(n, z - 1) + \sum_{i=1}^{q} f_\rho'(n, z, i) & n \neq q \\ \overline{f_\rho}(n, z) & n = q \end{cases} \quad (49)$$

As shown, we consider two forms form this case. The first form is when we have $n = q$, it means that all pixels can have $\pm z$ change. In this case, we have $f_\rho(n, z, q) = \overline{f_\rho}(n, z)$.

$$\overline{f_\rho}(n, z) = \overline{f_\mathcal{M}}(q, z). \overline{f_\rho}(n - q, z) + \overline{f_\mathcal{M}}(n - q, z). \overline{f_\rho}(q, z) \quad (50)$$

where, $\overline{f_\mathcal{M}}(n, z)$ is equal to $(2z + 1)^n$. In fact, to achieve the sum of changes when all the pixels can be changed, the pixels are divided as in two groups. This dividing can be numerical. For example, it can be converted into two groups of $1, n - 1$. Here, based on the $q$ value, the pixel group is divided into $q, n - q$. Thus, the sum of changes is equal to the number of the states of the first group in the sum of the changes of the second group plus the number of the states of the second group in the sum of the first group.

The second form is when $n \neq q$. In this case, we convert $f_\rho(n, z, q)$ into two separate sections and add them together. First section is dedicated to the states in which all pixels can have $\pm(z - 1)$ changes. Thus, the number of changes in this state is equal to $\overline{f_\rho}(n, z - 1)$. Second section is when we have $\pm(z)$ change. It is worth to mention that the number of pixels with $\pm(z)$ change can range from 1 to $q$. Thus, the sum of the amount of changes in this section is equal to the sum of the changes of different states as $\sum_{i=1}^{q} f_\rho'(n, z, i)$ is defined as follows. Later, each section is explained.

$$f_\rho'(n, z, q) = 2^q \binom{n}{q} . [q. z. \overline{f_\mathcal{M}}(n - q, z - 1) + \overline{f_\rho}(n - q, z - 1)] \quad (51)$$

**First section of relation.** The function $f_\rho'(n, z, q)$ is used when we have exactly $q$ pixels with $z$ changes. Indeed, exactly means exactly the sum of the changes of states are calculates that have $\pm z$ changes (The states without $\pm z$ are not considered). The states with $\pm z$ changes are converted into two sections. The first section is related to the pixels with $z$ changes and $q$ numbers. Thus, the amount of their changes is $q \times z$. The next section is related to the pixels with less than $z$ changes. Thus, we have n-q pixels that all have changes of $-(z - 1)$ to $+(z - 1)$. Thus, $\overline{f_\rho}(.)$ is used and $\overline{f_\rho}(n - q, z - 1)$ calculates the sum of changes in this section.

**Second section of relation.** As we have q pixel of $\pm z$ changes. Thus, of each state of $\pm z$ changes, we have a positive state and a negative state for each pixel as equal to $2^q$.

Now we have to choose the position of $q$ pixel among these $n$ pixels, thus, we use $\binom{n}{q}$. Finally, this number should be multiplied by the total of the two subsections above (e.g., $q$ pixels with $\pm z\ change$ and $n - q$ pixel with less than $\pm z$ changes). As mentioned, we achieve the relation $f_\rho'(n, z, q) = 2^q \binom{n}{q} . [q. z + \overline{f_\rho}(n - q, z - 1)]$. This relationship still is not complete. In fact, when we have $q$ pixels with $\pm z$, change, thought we have already calculated the position of this number $(q)$, its positive and negative, we have not considered the position and order of $n - q$ of other pixels in this section. Indeed, we should consider the number of states in which the sum of changes is equal but they have created different states based on the position and order of the pixels without any $\pm z$ change (e.g., we have three states with the sum of changes of 5, so, we have 15 changes in total). Finally, we should add multiplication by the number of the states to the relation. Finally, we reach the $q.z.\overline{f_\mathcal{M}}(n-q, z-1)$ in the original formula.

Finally, this relation is continued as the number of pixels in the group is equal to one. Thus, when in this function we have $n = q = 1$, it means we have one pixel ranging from $-z$ to $+z$. Thus, we can just add the changes together.

$$\overline{f_\rho}(1,z) = \sum_{i=-z}^{z} i = 2\sum_{i=0}^{z} i \quad (52)$$

Now, if we have $z = 0$ in this function, we have:

$$\overline{f_\rho}(n, 0) = 0 \quad (53)$$

It is clear that if in the original relation $f_\rho(n, z, q)$, $q = 0$, we have:

$$f_\rho(n, z, 0) = \overline{f_\rho}(n, z-1) \quad (53)$$

Also, for the proposed formula of embedding efficiency, as we need the square of the changes, the recurrence relation (51) is changed to (55) and relation (52) is changed to (56):

$$f_\rho'(n, z, q) = 2^q \binom{n}{q} . [q.z^2.\overline{f_\mathcal{M}}(n-q, z-1) + \overline{f_\rho}(n-q, z-1)] \quad (55)$$

$$\overline{f_\rho}(1,z) = 2\sum_{i=0}^{z} i^2 \quad (56)$$

## 7. Conclusion

In this study, after evaluating EMD and the relevant methods, a new formula was presented for better evaluation of the embedding efficiency in a specific related payload. Then, in order to better compare the methods, a chart of embedding efficiency (with the proposed relation and the existing relation) to the inverse related payload was plotted. For better comparison of the methods in these charts, we estimated an upper bound. This bound determines how much the best efficiency can be in any specified related payload. Hence, the methods that are closer to the bound provide better embedding efficiency in a certain related payload. Based on this definition, another criterion is introduced that is based on the distance of each method from the bound in the embedding efficiency diagram to the inverse related payload. Thus, the purpose of this research besides comparing EMD-based methods is to introduce an evaluation criterion to simultaneously measure the security and related payload of these techniques.